\documentclass[aps,prl,10pt,twocolumn,showpacs,superscriptaddress]{revtex4-2}
\usepackage{amssymb}
\usepackage{eurosym}
\usepackage{dcolumn}
\usepackage{bm}
\usepackage{graphicx}
\usepackage[dvipsnames]{xcolor}
\usepackage{amsmath}
\usepackage{amsfonts}
\usepackage{lmodern}
\usepackage{ulem}
\usepackage[colorlinks,linkcolor=black,citecolor=black,filecolor=black,urlcolor=Blue,bookmarksopen=false]{hyperref}
\usepackage{lipsum}

\begin{document}

\title{Atacamite Cu$_2$Cl(OH)$_3$ in High Magnetic Fields: Quantum Criticality and \\ Dimensional Reduction of a Sawtooth-Chain Compound}
\author{L. Heinze}
\email[Corresponding author:~]{l.heinze@fz-juelich.de}
\affiliation{Institut f\"{u}r Physik der Kondensierten Materie, Technische Universit\"{a}t Braunschweig, 38106 Braunschweig, Germany}
\affiliation{J\"{u}lich Center for Neutron Science (JCNS) at Heinz Maier-Leibnitz Zentrum (MLZ), Forschungszentrum J\"ulich GmbH, 85748 Garching, Germany}
\author{T. Kotte}
\affiliation{Hochfeld-Magnetlabor Dresden (HLD-EMFL) and W\"{u}rzburg-Dresden Cluster of Excellence ct.qmat, Helmholtz-Zentrum Dresden-Rossendorf, 01328 Dresden, Germany}
\author{R. Rausch}
\affiliation{Institut f\"{u}r Mathematische Physik, Technische Universit\"{a}t Braunschweig, 38106 Braunschweig, Germany}
\author{A. Demuer}
\affiliation{Universit\'{e} Grenoble Alpes, INSA Toulouse, Université Toulouse Paul Sabatier, CNRS, LNCMI, 38000 Grenoble, France}
\author{S. Luther}
\affiliation{Hochfeld-Magnetlabor Dresden (HLD-EMFL) and W\"{u}rzburg-Dresden Cluster of Excellence ct.qmat, Helmholtz-Zentrum Dresden-Rossendorf, 01328 Dresden, Germany}
\author{R. Feyerherm}
\affiliation{Helmholtz-Zentrum Berlin f\"{u}r Materialien und Energie, 14109 Berlin, Germany}
\author{E. L. Q. N. Ammerlaan}
\affiliation{High Field Magnet Laboratory (HFML-EMFL), Radboud University, 6525 ED Nijmegen, The Netherlands}
\author{U. Zeitler}
\affiliation{High Field Magnet Laboratory (HFML-EMFL), Radboud University, 6525 ED Nijmegen, The Netherlands}
\author{D. I. Gorbunov}
\affiliation{Hochfeld-Magnetlabor Dresden (HLD-EMFL) and W\"{u}rzburg-Dresden Cluster of Excellence ct.qmat, Helmholtz-Zentrum Dresden-Rossendorf, 01328 Dresden, Germany}
\author{M. Uhlarz}
\affiliation{Hochfeld-Magnetlabor Dresden (HLD-EMFL) and W\"{u}rzburg-Dresden Cluster of Excellence ct.qmat, Helmholtz-Zentrum Dresden-Rossendorf, 01328 Dresden, Germany}
\author{K. C. Rule}
\affiliation{Australian Nuclear Science and Technology Organisation, Lucas Heights, NSW 2234, Australia}
\author{A. U. B. Wolter}
\affiliation{Leibniz Institute for Solid State and Materials Research IFW Dresden, 01069 Dresden, Germany}
\author{H. K\"{u}hne}
\affiliation{Hochfeld-Magnetlabor Dresden (HLD-EMFL) and W\"{u}rzburg-Dresden Cluster of Excellence ct.qmat, Helmholtz-Zentrum Dresden-Rossendorf, 01328 Dresden, Germany}
\author{J. Wosnitza}
\affiliation{Hochfeld-Magnetlabor Dresden (HLD-EMFL) and W\"{u}rzburg-Dresden Cluster of Excellence ct.qmat, Helmholtz-Zentrum Dresden-Rossendorf, 01328 Dresden, Germany}
\affiliation{Institut f\"ur Festk\"orper- und Materialphysik, Technische Universit\"at Dresden Dresden, 01062 Dresden, Germany}
\author{C. Karrasch}
\affiliation{Institut f\"{u}r Mathematische Physik, Technische Universit\"{a}t Braunschweig, 38106 Braunschweig, Germany}
\author{S. S\"{u}llow}
\affiliation{Institut f\"{u}r Physik der Kondensierten Materie, Technische Universit\"{a}t Braunschweig, 38106 Braunschweig, Germany}
\date{\today}

\begin{abstract} 
We report an extensive high-field study of atacamite Cu$_2$Cl(OH)$_3$, a material realization of quantum sawtooth chains with weak interchain couplings, in continuous and pulsed magnetic fields up to 58\,T. In particular, we have mapped the entropy landscape for fields as high as 35\,T and have identified a field-induced quantum critical point at $21.9(1)$\,T for $\mathbf{H} \parallel c$ axis. The quantum critical point separates field regions with and without magnetic order, evidenced by our thermodynamic study and $^1$H nuclear magnetic resonance spectroscopy, but lies far below full saturation of the magnetization. Corroborated by numerical results using density-matrix renormalization group (DMRG) calculations, we find this behavior associated with a dimensional reduction of the spin system: the sawtooth chain effectively decouples into an antiferromagnetic spin-$1/2$ chain (backbone of the sawtooth chain) in the presence of an exchange field produced by the remaining field-polarized spins.
\end{abstract} 

\maketitle

Frustrated quantum magnets are inherently characterized by highly degenerate magnetic ground states~\cite{Balents2010,Broholm2020}. These are susceptible to tuning by applied pressure or external magnetic field which may result in a change of effective dimensionality of the underlying spin system. Such tuning can lead to quantum phase transitions from an antiferromagnet to a quantum paramagnet or vice versa~\cite{Sebastian2006,Roesch2007,Skoulatos2017,Hirai2019}. The emergent quantum critical points (QCPs) induced by such ground-state tuning are outstanding points in the corresponding phase diagrams: They are accompanied by highly structured entropy landscapes and quantum-critical scaling behavior of thermodynamic quantities~\cite{Zhu2003,Garst2005,Vojta2018}. In the case of a field-induced QCP, large temperature changes during an adiabatic magnetization can occur~\cite{ZhitomirskyHonecker2004,Honecker2009}. This makes frustrated quantum magnets candidates for application in modern-day low-temperature cooling devices~\cite{Lang2013,Tokiwa2021,Arjun2023a,Arjun2023b,Reichert2023}. At present, however, there is a substantial lack of knowledge in particular about the phenomenon of dimensional reduction in frustrated magnetism as an essential element of quantum criticality.

An additional aspect comes into play for material realizations of low-dimensional quantum magnets. The anisotropic orbital arrangements of magnetic and nonmagnetic ions in a crystal can lead to a magnetic system with a coupling motif of effectively lower dimensionality than the three-dimensional (3D) crystal structure. For such low-dimensional magnets, the magnetic behavior on a low energy scale is influenced by weak residual exchange interactions, and long-range magnetic order might result at low temperatures (see, for instance, Refs.~\cite{Banks2009,Rule2011}). This clearly distinguishes real magnets from their purely low-dimensional model counterparts studied theoretically. In this situation, in order to firmly establish quantum criticality and changes of the effective dimensionality of a spin system in real magnets, an extensive set of macroscopic and microscopic experiments as well as accompanying theoretical work is required. Here, we present such a combined study revealing a unique quantum spin behavior in a sawtooth-chain compound associated with the occurrence of a QCP. In our case, we consider it a special type of dimensional reduction when the energy scale of the 3D residual couplings is overcome.

\begin{figure}[t!]
\centering
\includegraphics[width=\linewidth]{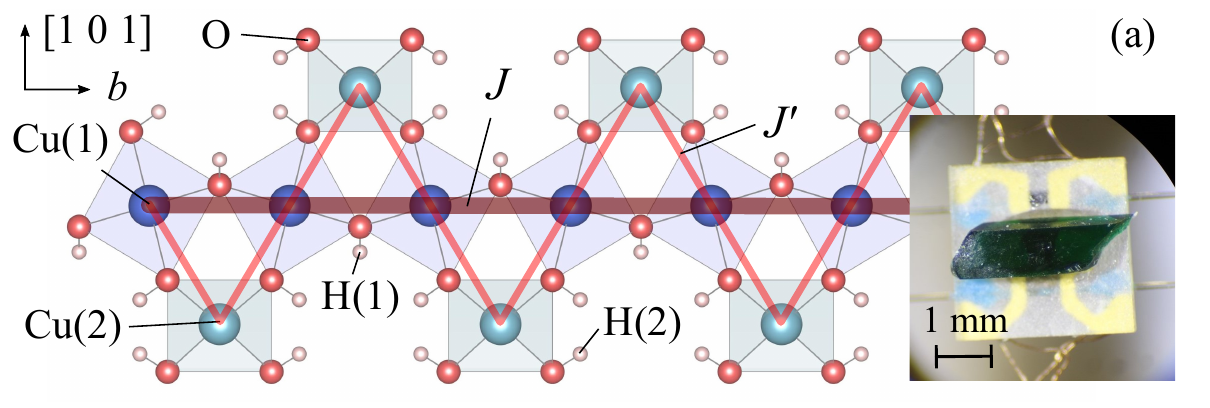}
\includegraphics[width=\linewidth]{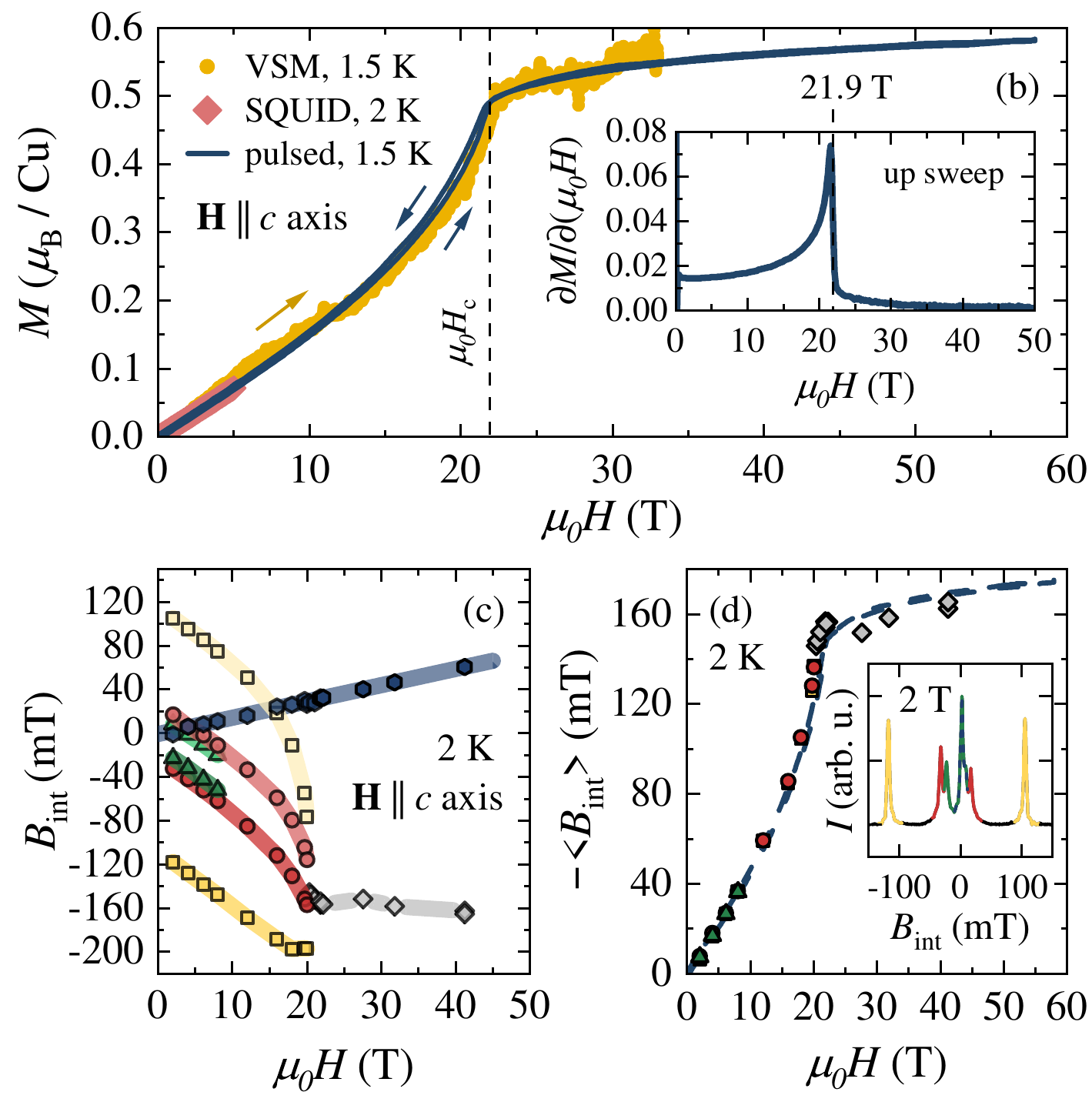}
\caption{(a) Sawtooth-chain unit in atacamite (space group $Pnma$) formed by Cu(1) (basal) and Cu(2) (apical) sites; structural data from Ref.~\cite{Parise1986}. The Cu-O plaquettes are completed as Cu(1)(OH)$_4$Cl$_2$ and Cu(2)(OH)$_5$Cl octahedra~\cite{SI}. Inset: Atacamite single crystal on a puck for heat-capacity measurements. (b) High-field magnetization for $\mathbf{H} \parallel c$ axis measured using a vibrating sample magnetometer (VSM) in DC fields as well as pulsed-field magnetometry~\cite{SI}. We used SQUID data for scaling the pulsed-field data. Inset: Field-derivative of the magnetization (pulsed). (c)--(d) $^1$H-NMR results for $\mathbf{H} \parallel c$ axis. A typical $^1$H-NMR spectrum for the low-field AFM ground state is shown in the inset of (d). We plot the field dependence of (c) the internal fields $B_\mathrm{int}$ of the individual resonance lines, as well as (d) the mean shift of each line pair $\langle B_{\text{int}} \rangle$, which represents the local uniform magnetization within the AFM phase. The blue dashed line in (d) corresponds to the bulk magnetization from panel (b) scaled by $-0.33\,\mathrm{T}\,\mu_\mathrm{B}^{-1}$.}
\label{fig:Fig1}
\end{figure}

As object of study we choose the antiferromagnetic (AFM) sawtooth chain, or delta chain, a fundamental model in one-dimensional (1D) frustrated quantum magnetism, which has been studied theoretically for decades~\cite{Kubo1993,Schulenburg2002,Blundell2003,ZhitomirskyHonecker2004,Zhitomirsky2004,Richter2004,Derzhko2015,Metavitsiadis2020,Richter2022}. It is described by the Hamiltonian

\begin{align}
\mathcal{H}=&\sum_{i~\mathrm{odd}}J\mathbf{S}_{i}\cdot\mathbf{S}_{i+2}+J'\big(\mathbf{S}_{i}\cdot\mathbf{S}_{i+1}+\mathbf{S}_{i+1}\cdot\mathbf{S}_{i+2}\big) \nonumber \\
& -h\sum_i {S}^z_{i}
\label{eq:Hamiltonian_sawtoothchain}
\end{align}

\noindent where $\mathbf{S}_{i}$ is the spin-$1/2$ operator at site $i$. The exchange interaction $J$ couples the neighboring spins along the basal line of the chain while $J'$ is the exchange interaction between basal ($i$ odd) and apical spins [$i$ even; Fig. \ref{fig:Fig1}(a)]. The term $-h\sum_i {S}^z_{i}$ accounts for the presence of a magnetic field with Zeeman energy $h = g\mu_\mathrm{B}\mu_0H$. Most notably, the existence of flat magnon bands in the energy spectrum of this model led to the prediction of a number of exotic in-field phenomena such as magnetization plateaus, macroscopic magnetization jumps or novel magnetoelectric effects~\cite{Schulenburg2002,Zhitomirsky2004,Derzhko2015,Richter2022}. As well, a pronounced magnetocaloric effect (MCE) around the field of full field polarization has been predicted by theory~\cite{ZhitomirskyHonecker2004}.

In this Letter, we present the unique case of the frustrated quantum spin system atacamite Cu$_2$Cl(OH)$_3$. We recently established this compound as a sawtooth-chain material with small coupling ratio $J^\prime/J \sim 1/3$ and weak residual interchain couplings, which lead to AFM long-range order below $T_\mathrm{N} = 8.9(1)$\,K~\cite{Heinze2021,comm:T_N}. Using thermodynamic as well as microscopic probes under high magnetic fields up to 58\,T, we demonstrate here that for this compound an applied magnetic field serves as control parameter to tune the system through a QCP at $\mu_0H_\mathrm{c} = 21.9(1)$\,T ($\mathbf{H} \parallel c$ axis). Associated with this quantum phase transition, we find a reduction of the effective dimensionality of the spin system together with a suppression of long-range magnetic order close to half saturation of the magnetization. This finding is corroborated by numerical results, which show that close to half saturation for the \textit{bare} quantum sawtooth chain with $J^\prime/J = 0.3$ (bare: no interchain couplings), a decoupling of the basal and apical spins into two independent subsystems occurs. Here, the high-field correlations of the basal spins closely resemble those of the spin-$1/2$ AFM Heisenberg chain. For atacamite, at the QCP the weakly coupled 3D AFM state is replaced by a field-induced quantum spin liquid (QSL) consisting of 1D subunits formed by the basal spins in the spin-polarized background of the apical spins.

Atacamite~\cite{Parise1986}, a dark green mineral, was recently identified as material realization of sawtooth chains of Cu spins [$J = 336$\,K, $J^\prime = 102$\,K; Fig. \ref{fig:Fig1}(a)] in a weak 3D network of interchain couplings~\cite{Heinze2021,Heinze2018}. Experimentally, this material stands out through its unusual high-field magnetization, closely resembling that of a wide $1/2$-plateau above a field of 31.5\,T for $\mathbf{H} \parallel b$ axis~\cite{Heinze2021}. This magnetization behavior, however, was found to be unrelated to the $1/2$-plateau of the quantum sawtooth chain~\cite{Richter2004}. Here, we will provide evidence that the flattening of the magnetization coincides with the suppression of AFM order with the aforementioned QCP being the low-$T$ endpoint of the AFM phase boundary.

A pronounced flattening of the high-field magnetization $M(\mu_0H)$ is present also for the experimental geometry $\mathbf{H} \parallel c$ axis, verified using both pulsed and slowly sweeped external magnetic fields [Fig. \ref{fig:Fig1}(b)]~\cite{Skourski2011,SI}. Again, the flattening occurs at $M \sim M_\mathrm{sat}/2$. However, for $\mathbf{H} \parallel c$ axis, the magnetization flattens already at 21.9\,T [inset Fig. \ref{fig:Fig1}(b); critical field determined as the maximal drop of the field derivative of $M(\mu_0H)$]. This leads to a much easier experimental accessibility of the high-field region, especially in static magnetic fields.

By means of $^1$H nuclear magnetic resonance (NMR) spectroscopy in static as well as pulsed fields up to $41$\,T, we probe the local hyperfine interaction acting on the hydrogen nuclear moments. This provides microscopic insights into the magnetic order and its suppression in atacamite [Fig. \ref{fig:Fig1}(c)--(d)]. According to the crystal symmetry, we expect two $^1$H lines in the paramagnetic state. At 2\,K, below $T_\mathrm{N}$, the ground-state NMR spectrum of atacamite reveals seven resonance frequencies ($f_\mathrm{res}$). Six resonances are attributed to the H(2) nuclear positions [inset Fig. \ref{fig:Fig1}(d)], while the local symmetry in atacamite prevents further line splitting of the H(1) line [blue hexagons in Fig. \ref{fig:Fig1}(c)] in the studied field geometry.

Focusing on the H(2) lines, the corresponding internal fields, $B_\mathrm{int} = 2\pi/\gamma \times f_\mathrm{res} - \mu_0H$ ($\gamma / 2\pi = 42.5774$\,MHz/T), are pairwise symmetric to zero at low fields, indicating AFM order. We present in Fig. \ref{fig:Fig1}(c) the field dependence of the internal fields $B_\mathrm{int}$ obtained for all observed $^1$H resonances. Additionally, we plot the mean shift of each line pair, $\langle B_\mathrm{int} \rangle$, in Fig. \ref{fig:Fig1}(d). We find that $\langle B_\mathrm{int} \rangle$ scales with the magnetization across the entire field range for all pairs [blue dashed line in Fig. \ref{fig:Fig1}(d)], confirming the consistency of  the $^1$H-NMR and magnetization data. The splitting of the symmetric NMR lines [Fig. \ref{fig:Fig1}(c)] changes only marginally with field below 16\,T. Above 8\,T, the two inner pairs cannot be resolved. As the field increases further, the NMR line splitting decreases significantly until all lines merge above 20.3\,T, indicating the vanishing of the staggered magnetization. Thus, the $^1$H-NMR results evidence that the flattening of the magnetization curve at $M \sim M_\mathrm{sat}/2$ coincides with the suppression of long-range AFM order.

\begin{figure}[t!]
\centering
\includegraphics[width=\columnwidth]{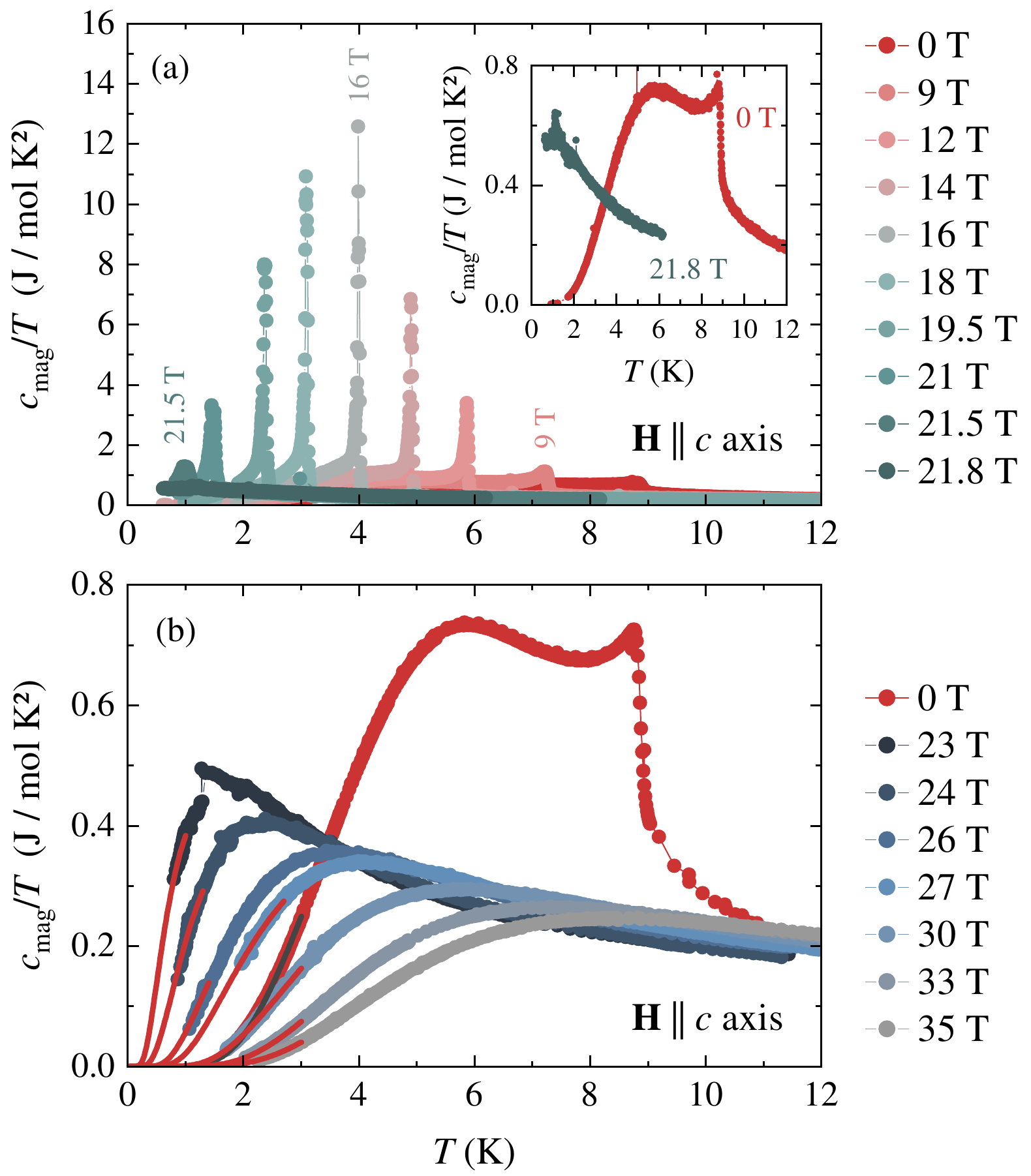}
\caption{(a)--(b) $c_\mathrm{mag}/T$ of atacamite measured in fields up to 35\,T ($\mathbf{H} \parallel c$ axis). The inset in (a) shows a zoom-in on the scale of panel (b). Exemplary fits to the low-$T$ data are shown as dark gray (0\,T) and red ($\mu_0H \geq 23$\,T) curves in (b).}
\label{fig:Fig2}
\vspace{3mm}
\includegraphics[width=\columnwidth]{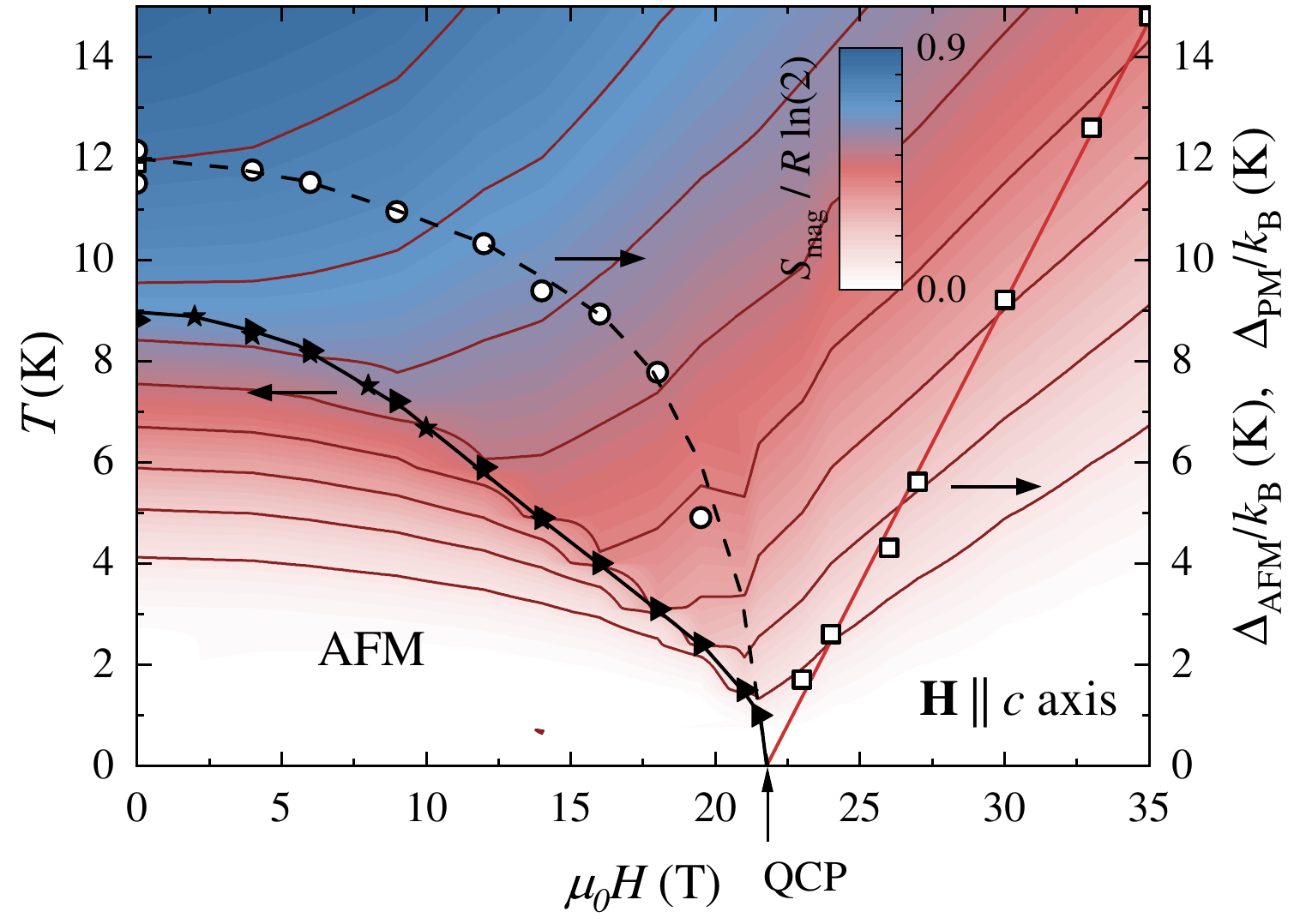}
\caption{Entropy landscape $S_\mathrm{mag}(T, \mu_0H)$ ($\mathbf{H} \parallel c$ axis). Dark red curves on the contour are interpolated isentropes with an increment of $0.1\,R\,\ln(2)$. Black symbols denote $T_\mathrm{N}$ determined from heat capacity (triangles) and magnetization (stars). White symbols show $\Delta_\mathrm{AFM}$ (circles) and $\Delta_\mathrm{PM}$ (squares)~\cite{SI}. We fitted $\Delta_\mathrm{PM}$ linearly (red line); see text.}
\label{fig:Fig3}
\end{figure}

Next, we investigate the thermodynamic properties of atacamite in the two field regions with and without long-range AFM order by carrying out heat-capacity measurements in static magnetic fields up to 35\,T ($\mathbf{H} \parallel c$ axis) and down to $^3$He temperatures. The behavior of the magnetic contribution to the specific heat $c_\mathrm{mag}$~\cite{SI}, plotted as $c_\mathrm{mag}/T$ (Fig. \ref{fig:Fig2}), can clearly be distinguished into the field region of the AFM phase and that beyond the AFM phase. In the first region below $21.8$\,T [Fig. \ref{fig:Fig2}(a)], we observe the transition anomaly at $T_\mathrm{N}$ which shifts towards lower temperatures in applied fields. The field evolution of $T_\mathrm{N}$ is equally observed in the temperature-dependent magnetization~\cite{SI}. Furthermore, in intermediate fields, we observe an enormous sharpening of the $c_\mathrm{mag}$-anomaly at $T_\mathrm{N}$, now resembling a first-order phase transition. The maximal $c_\mathrm{mag}/T$ in 16\,T is close to 13\,J\,mol$^{-1}$\,K$^{-2}$. For higher fields, the anomaly weakens again. From the heat-capacity results, the phase boundary can be traced down to (1\,K, 21.5\,T), leading to the conclusion that AFM order is fully suppressed at $\mu_0H_\mathrm{c} = 21.9(1)$\,T. In the second field region above 23\,T [Fig. \ref{fig:Fig2}(b)], the transition anomaly of $c_\mathrm{mag}/T$ vanishes and a broad anomaly appears~\cite{comm:23Tcurve}. It is reminiscent of a behavior produced by low-lying spin excitations with a field-dependent energy gap $\Delta_\mathrm{PM}$.

To establish the entropic landscape in the field-temperature phase diagram of atacamite, we evaluate the magnetic entropy $S_\mathrm{mag}$ by integration of the specific-heat data as $S_\mathrm{mag}(T) = \int_0^T c_\mathrm{mag}(T^\prime)/T^\prime dT^\prime$~\cite{SI}. In Fig. \ref{fig:Fig3}, we present a contour plot of $S_\mathrm{mag}(T,\mu_0H)$, thereby linearly interpolating between the entropy data calculated for a fixed field. We added the $T_\mathrm{N}$ values extracted from the maximum of $c_\mathrm{mag}/T$ as well as the maximum of $\partial M/\partial T$ data~\cite{SI} to Fig. \ref{fig:Fig3}. They denote the AFM phase boundary. We note that the change of the AFM transition towards a first-order character in applied field [Fig.~\ref{fig:Fig2}(a)] might reflect an additional complexity of the in-field behavior which is of interest for future work.

The entropy landscape in Fig. \ref{fig:Fig3} is highly distorted giving rise to a pronounced MCE upon an adiabatic (de)magnetization~\cite{SI}. The isentropes run along decreasing temperature when increasing the magnetic field within the AFM phase and along increasing temperature when increasing the magnetic field beyond the AFM phase. A sign change of $\left({\partial T}/{\partial (\mu_0 H)}\right)_{S_\mathrm{mag}}$ occurs at or very close to the AFM phase boundary. This behavior reflects the presence of a field-induced QCP~\cite{Zhu2003,Garst2005,Vojta2018}.

Associated with the occurrence of the QCP at 21.9(1)\,T is an anomalous evolution of the low-energy spin excitations reflected by the thermodynamic quantities. Below 21.9\,T, the suppression of AFM order is reflected by the suppression of a magnon gap $\Delta_\mathrm{AFM}$ in the magnetic insulator atacamite, which we model by fitting the low-temperature specific heat with $c_\mathrm{mag} \propto \exp\left(-\Delta_\mathrm{AFM}/k_\mathrm{B}T \right)$ [exemplary dark gray fit curve for 0\,T in Fig. \ref{fig:Fig2}(b)]. Beyond the field-induced suppression of AFM order, the low-temperature specific heat can again be fitted with $c_\mathrm{mag} \propto \exp\left(-\Delta_\mathrm{PM}/k_\mathrm{B}T \right)$ [exemplary red fit curves in Fig. \ref{fig:Fig2}(b)]. The field-dependent gap parameters $\Delta_\mathrm{AFM}$ and $\Delta_\mathrm{PM}$ determined this way are included in Fig. \ref{fig:Fig3}. Overall, the field evolution of $\Delta_\mathrm{AFM}$ and $\Delta_\mathrm{PM}$ reflects a field-driven closing ($\Delta_\mathrm{AFM}$) and reopening ($\Delta_\mathrm{PM}$) of the spin excitation gaps in atacamite at the QCP. Notably, the field evolution of $\Delta_\mathrm{PM}(\mu_0H)$ appears to evolve as $\Delta_\mathrm{PM}/k_\mathrm{B} = A\,(\mu_0H-\mu_0H_\mathrm{c})$. A corresponding fit yields $A = 1.11(3)$\,K/T and $\mu_0H_\mathrm{c} = 21.8(2)$\,T [red line in Fig. \ref{fig:Fig3}], in full agreement with the field at which the magnetization flattens and AFM order is suppressed.

To account for the emergence of the QCP, we propose the following scenario: The dominant exchange interaction $J$ is between the basal spins [Cu(1)], which form strong antiferromagnetic bonds, while the apical spins [Cu(2)] are only weakly coupled to them by $J^\prime$. Frustration inhibits a clear alignment of the apical spins with respect to the basal ones and they become more easily polarizable~\cite{Rausch2024}. However, since they also mediate the 3D coupling between the basal chains~\cite{Heinze2021}, this coupling is broken once the apical spins become fully aligned in an external field. The behavior beyond the QCP is then characterized by an ensemble of decoupled spin-$1/2$ AFM Heisenberg chains, formed by the basal spins, in the presence of a field-polarized ensemble of the apical spins.

\begin{figure}[t!]
\begin{center}
\includegraphics[width=\columnwidth]{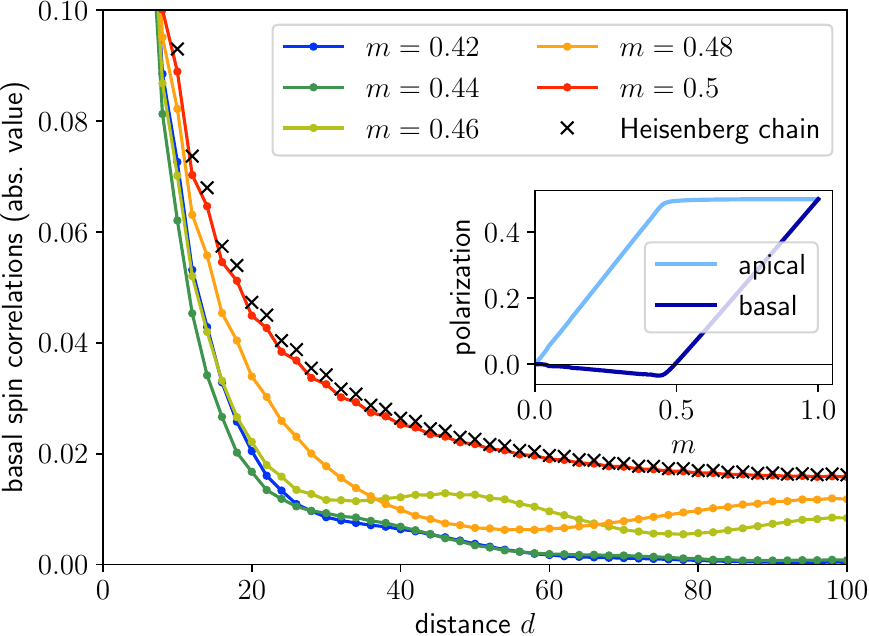}
\caption{Basal spin correlations $\left|\left<\mathbf{S}_i\cdot\mathbf{S}_{i+d}\right>-\left<\mathbf{S}_i\right>\cdot\left<\mathbf{S}_{i+d}\right>\right|$ ($i$ odd, $d$ even) for a quantum sawtooth chain of $L=200$ sites (100 unit cells, PBC), $J=1$, $J'=0.3$, in various sectors of the total spin $m=S_{\text{tot}}/(L/2)$ up to half saturation $m=0.5$. The black crosses indicate the values for a pure Heisenberg chain $\mathcal{H}=J\sum_i\mathbf{S}_i\cdot\mathbf{S}_{i+1}$, with the distance $d$ scaled accordingly. Inset: The corresponding polarization $\sum_i\left<S^z_i\right>$ for the apical ($i$ even) and basal ($i$ odd) sites.}
\label{fig:Fig4}
\end{center}
\end{figure}

We present numerical evidence for this scenario by performing density-matrix renormalization group (DMRG) calculations for $L=200$ sites with periodic boundary conditions (PBC)~\cite{Rausch2024}. Figure \ref{fig:Fig4} shows the spin-spin correlations of the basal spins in different sectors of the total spin up to half saturation. We observe an increasing correlation length with $m$, and at the point of half saturation the correlations are well-described by a pure spin-$1/2$ AFM Heisenberg chain. The inset of Fig. \ref{fig:Fig4} shows how the polarization is distributed among the apical and basal sites. For small $m$, the apical spins are strongly polarized along the field, while the basal spins are at first weakly polarized in the opposite direction to optimize their mutual exchange interaction. The polarization of the basal spins passes through zero at half saturation. At the same time, the apical spins become fully saturated, and the two subsystems decouple from each other. A more detailed analysis should explicitly demonstrate the creation of the gap via staggered fields resulting from 3D couplings. We leave this for future work.

Experimentally, we find support for this scenario in the following observations: (1) The AFM long-range order in atacamite is fully suppressed at the critical field of 21.9(1)\,T [evidenced by NMR; Fig. \ref{fig:Fig1} (c)] although this field lies far below the expected field of full field polarization [Fig. \ref{fig:Fig1}(b)]. This absence of long-range order indicates an effective dimensional reduction of the spin system from a 3D magnetically ordered phase to an effectively one-dimensional behavior. The related QCP, low-$T$ endpoint of the AFM phase boundary, stands out from a highly distorted entropy landscape. (2) At magnetic fields beyond the AFM phase, the internal fields obtained by means of $^1$H-NMR closely resemble the calculated ones within a scenario of a field-polarized Cu(2) subsystem (c.f. Supplemental Material~\cite{SI}). This is further supported by a macroscopic magnetization of $M \sim M_\mathrm{sat}/2$ in this field region. (3) The low-temperature behavior of the heat capacity of atacamite for $\mu_0H > \mu_0H_\mathrm{c}$ reflects a field-driven opening of a spin excitation gap $\Delta_\mathrm{PM}$ in accordance with full field polarization of the Cu(2) magnetic subsystem (apical spins). The linear behavior of $\Delta_\mathrm{PM}(\mu_0H)$ indicates the field-growing energy difference between fully field-polarized apical spins and the first excitation of this spin system which is an antiferromagnetic excitation. In the same field region, the remaining Cu(1) subsystem (basal spins) forming a weakly field-polarized AFM spin-$1/2$ chain with $J = 336$\,K~\cite{Heinze2021} does not contribute significantly to the low-temperature heat capacity of atacamite~\cite{SI,Johnston2000}. (4) The entropy landscape is highly distorted, giving rise to a large MCE, and the isentropes close to the QCP behave different in the AFM region and beyond. While we cannot conclude on the dimensionality of the spin system from $T(\mu_0H)_{S_\mathrm{mag}}$, we note that this reflects a change of the quantum critical exponents and/or the dimension from the theoretical descriptions of quantum phase transitions~\cite{Garst2005}.

In summary, the mineral atacamite is a unique example to study the phenomenon of dimensional reduction at a QCP. Through extensive microscopic and macroscopic studies in high magnetic fields, and complemented by DMRG calculations, we find that with the suppression of AFM long-range order at the QCP, the behavior of a bare sawtooth chain is recovered via an effective dimensional reduction of the spin system. Beyond the fundamental implications of this new state, the strong impact of the QCP on the entropy map highlights the potential relevance of frustrated quantum magnets for low-temperature cooling applications.

\begin{acknowledgments}
We acknowledge the support of the HLD-HZDR, the LNCMI-CNRS, as well as the HFML-RU/NWO-I, members of the European Magnetic Field Laboratory (EMFL). This work was partially supported by the DFG under Contract No.~SU229/9-2. We acknowledge fruitful discussions with W. Brenig, T. Gottschall, A. Honecker, M. Jaime and M. Zhitomirsky. We thank G. Paskalis and J. McAllister for supplying us with atacamite crystals used for this study and thank C. Kleeberg for support with orienting the sample used for the heat capacity and NMR measurements at HLD-HZDR. We further thank T. Lampe for the support during the sample preparation, J. Grefe and D. Menzel for assistance with the SQUID measurements, as well as S. Findeisen for his support in implementing the two-axis rotation in the NMR experiments. Fig. \ref{fig:Fig1}(a) and Fig. S1 (in Ref.~\cite{SI}) were drawn using VESTA 3~\cite{VESTA}.
\end{acknowledgments}

L.H. and T.K. contributed equally to this work.


\begin{thebibliography}{999}
\bibitem{comm:23Tcurve} The step in the heat-capacity data recorded at 23\,T at $\sim 1$\,K is likely to be caused by a slightly drifting bath temperature during the measurement.
\end{thebibliography}
\end{document}